\documentclass[conference,10pt,a4paper]{IEEEtran}
\usepackage{graphicx}
\usepackage{amsthm}
\usepackage{empheq}
\usepackage{amsfonts}
\usepackage{amsmath,amsthm,amssymb}
\usepackage{graphicx,subfigure}
\usepackage{color}
\usepackage{txfonts}
\usepackage{booktabs,array}
\usepackage{multirow}
\usepackage{url}
\usepackage{tikz}
\usepackage{booktabs}
\usepackage[utf8]{inputenc}

\renewcommand{\eqref}[1]{(\ref{#1})}

\usepackage{epstopdf}
\usepackage{steinmetz}

\usepackage[caption=false,font=footnotesize,labelfont=sf,textfont=sf]{subfig}
\usepackage{flexisym}
\usepackage[T1]{fontenc}
\newcounter{MYtempeqncnt}
\begin{document}
	\title{Analysis of Key Generation Rate from Wireless Channel in In-Band Full-Duplex Communications}
	\author{\IEEEauthorblockN{Alireza Sadeghi$^{*}$, Michele Zorzi$^\#$, Farshad 
			Lahouti$^{+}$}
		
		\IEEEauthorblockA{$^*$Electrical and Computer Engineering Department, University of Minnesota, USA\\ }
		\IEEEauthorblockA{$^\#$Department of Information Engineering, University of Padova, Italy}
		\IEEEauthorblockA{$^+$Electrical Engineering Department, California 
		Institute of Technology, USA\\ }

		{sadeg012@umn.edu, zorzi@dei.unipd.it, lahouti@caltech.edu}}	
	
	\maketitle
	
\begin{abstract}
	In this paper, the impact of in-band full-duplex (IBFD) wireless communications on secret key generation via physical layer channel state information is investigated. A key generation strategy for IBFD wireless devices to increase the rate of generated secret keys over multipath fading channels is proposed. Conventionally, due to the half-duplex (HD) constraint on wireless transmissions, sensing simultaneous reciprocal channel measurements is not possible, which leads to a degraded key generation rate. However, with the advent of IBFD wireless devices, the legitimate nodes can sense the shared wireless link simultaneously at the possible cost of a self-interference (SI) channel estimation and some residual self-interference (RSI). As we demonstrate, with HD correlated observations the key rate is upper bounded by a constant, while with IBFD the key rate is only limited by the SI cancellation performance and is in general greater than that of its HD counterpart. Our analysis shows that with reasonable levels of SI cancellation, in the high SNR regime the key rate of IBFD is much higher, while in low SNRs, the HD system performs better. Finally, the key rate loss due to the overhead imposed by the SI channel estimation phase is discussed.

\end{abstract}

\IEEEpeerreviewmaketitle

	\section{Introduction}
	The broadcast nature of wireless transmission makes communications over such channels vulnerable to eavesdropping. The security of communications over these channels is traditionally ensured by cryptographic schemes. However, \textit{physical layer} security mechanisms can provide a strengthened level of security in communications \cite{Survey1}. In this paper, secret key generation based on physical channel state information (CSI) in point-to-point in-band full-duplex (IBFD or FD) wireless communications is investigated.

	The concept of \textit{secret key rate} for a system of two legitimate nodes sharing some common source of information (randomness) is developed in \cite{Maurer}, where upper and lower bounds on key generation rate are presented when a public and error free feedback channel is assumed available. In \cite{hassan1996cryptographic}, the characteristics of a reciprocal wireless communication channel between a pair of legitimate nodes are suggested as the source of common randomness. Since then, generating secrecy keys based on wireless channel characteristics has attracted much research attention.  Some of these works propose to generate keys relying on the phase of the fading channel \cite{PhaseSec}, differences of phases between two frequency tones \cite{SayeedMultitone}, level crossing in legitimate nodes \cite{levelcrossing}, direct quantization of the complex channel coefficients \cite{directlyquantizing}, and unknown deterministic parameters estimated by legitimate nodes \cite{deterministicestimation}. 
	
	In many of the works in this context, Time Division Duplexed (TDD) communication is considered. This is due to the fact that key generation relies on channel reciprocity, which is assumed valid in TDD scenarios. However, in rich scattering environments, due to reasons such as the spatial changes of the legitimate nodes or any scattering medium between them and also the temporal changes of the channel, the channel sensed over two consecutive time-slots is not identical. This reveals that although the channel is inherently reciprocal, in half-duplex (HD) systems the two legitimate nodes merely observe correlated channel characteristics (Fig. 1). To overcome this issue, the authors in \cite{HRUBE} introduced a fractional interpolation filtering framework to enable two HD legitimate nodes to find what the channel measurements would have been if the nodes had made simultaneous measurements by the help of a finite impulse response filtering. The fractional interpolation is applied to the real non-simultaneous channel measurements at the two nodes. The extension of \cite{HRUBE} in \cite{ARUBE} also takes into account the non-reciprocities caused by different hardware characteristics. Despite its low secret key generation rate, in \cite{Yener} and \cite{Infocom2012} the channel deep fades are proposed to facilitate a common source of randomness counting on the fact that such channel fades can be long enough to be sensed by both nodes.
	
	In this paper, different from the above works, we study the possible improvements in the secrecy key generation rate by simultaneously sensing the channel using IBFD wireless devices in the presence of residual self-interference (RSI). We present a scheme for the IBFD pair of nodes to obtain estimates of the states of their self-interference and direct channels, and then analyze the resulting secret key generation rate. In this analysis, we also take into account the non-reciprocity inherent in sequential wireless channel sensing. The results demonstrate that FD communications can noticeably boost the secret key rate in an operationally interesting regime of parameters. The prior works related to IBFD based secure communication have not dealt with secret key generation and are mainly focused on enhancing the secrecy performance of wireless communications. The effects of artificial noise generation and active pilot contamination attack on secrecy performance are studied in \cite{FDBasestation} and \cite{Maham}, respectively, whereas the effect of IBFD relays on the secrecy rate is investigated in \cite{SecrecyFriendlyFDrelayWCNC2015}. 
	
	The rest of this paper is organized as follows. Some basic prerequisites on IBFD are provided in Section \ref{Section:In-Band Full-Duplex Communication}. The system model in both HD and FD cases is described in Section \ref{Section:System Model}. The key rate analysis in both cases are discussed in Section \ref{Secrecy Key Generation}. Section \ref{Section: Simulation Results and Analysis} presents the simulation results and some analysis. Finally, Section \ref{Section: Conclusion} concludes the paper. 
	
	\begin{figure}[!t]
		\centering
		\includegraphics[width=0.45 \textwidth]{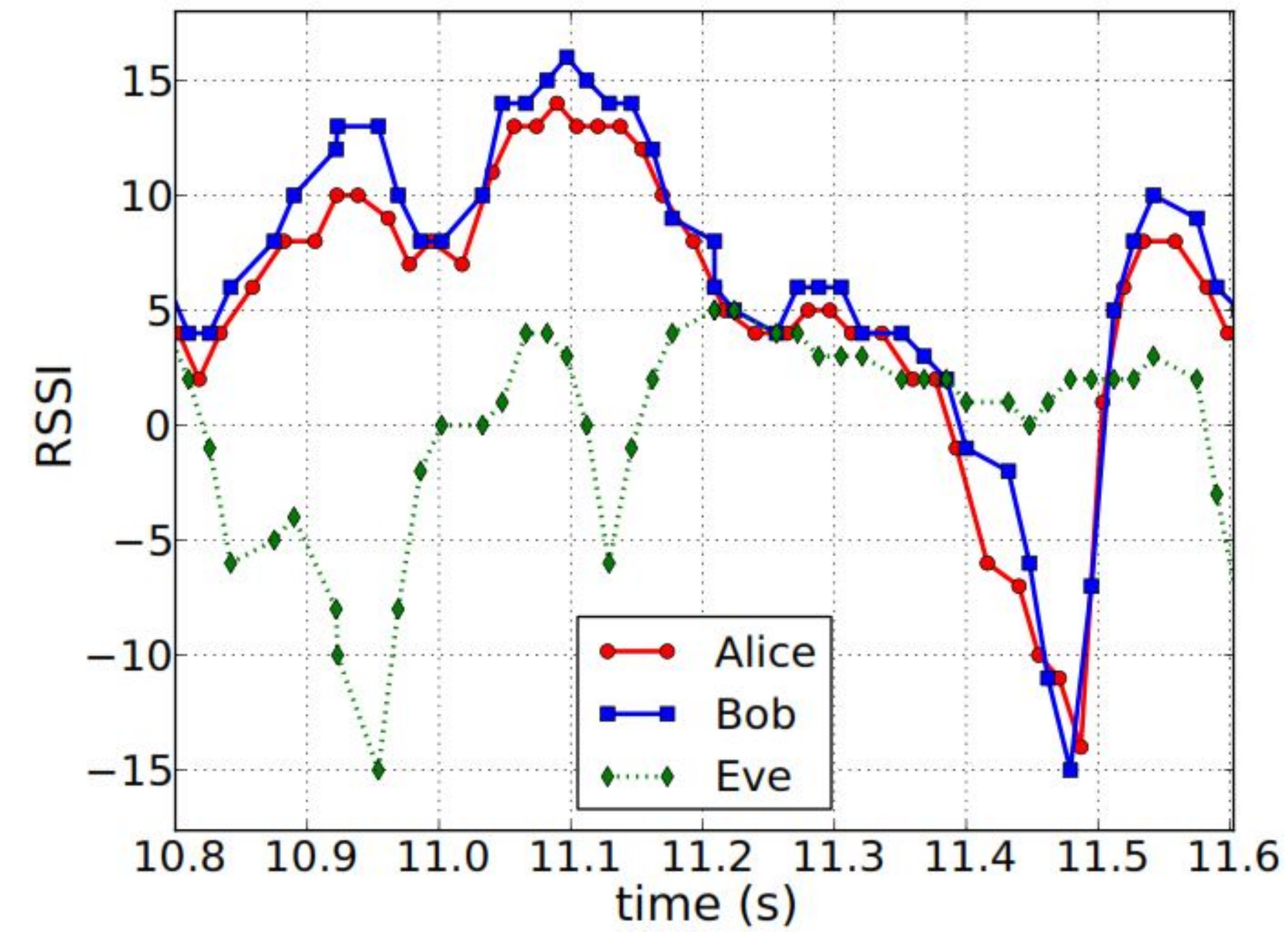}
		\caption{Received signal strength measurements taken over time. Alice and Bob's
			RSS measurements are not the same, but correlated \cite{TheFig.1}.}
		\label{Fig0}
	\end{figure}
	
	\section{Preliminary: In-Band Full-Duplex Communication}
	\label{Section:In-Band Full-Duplex Communication}
	In In-Band Full-Duplex (IBFD or FD), two nodes conduct a two-way communication over a single wireless channel simultaneously. 
	
	Conventionally, in an HD wireless device, the ratio of the self-interference (SI) generated by the ongoing transmission compared to that of the desired signal had made any FD communication infeasible. This high SI is simply due to the fact that the SI signal is traveling a much shorter distance compared to the desired one, leading the desired signal to be soaked in the SI. However, recently by the advent of different passive and active suppression mechanisms an FD wireless device can suppress its own SI strongly enough to enable simultaneous transmission and reception in a single frequency band at the same time. However in practice, because of the many imperfections in the transceiver operation, fully canceling SI is not possible, and therefore some residuals will remain, known as RSI. The RSI represents the main obstacle for a perfect FD communication, and is similar to noise and essentially uncorrelated with the original transmitted signal. 
	
	Among different methods to model the RSI, a zero mean Gaussian random variable is used in this paper as follows \cite{RSIModel} 
	\begin{equation}
		\text{RSI} \sim \mathbb{CN} \left( {0,\sigma_{RSI}^2} \right)
		\label{eq.RSI}
	\end{equation}
	where $\sigma_{RSI}^2$ is the variance of the RSI signal, and commonly is a function of the power of the ongoing transmitted signal.
	
	In the given time frame for generating a key in bidirectional FD operation that is provided in Fig. \ref{fig.2.b}.b, the FD transceivers need to allocate a portion of the overall time interval to estimate their own SI channel state information (CSI), calculate the coefficients of the analog SI cancellation circuits, and calculate the digital SI canceller parameters. This is the overhead to be paid in the system for making any FD operation possible \cite{DaurteExperimental}. We call this interval as the SI channel estimation phase. 
	
	\section{System Model}
	\label{Section:System Model}
	In our scenario two legitimate nodes, Alice and Bob, wish to establish a shared secure key based on their shared wireless fading channel in the presence of a passive eavesdropper, Eve. As usually done in the related literature, we assume that, in addition to the wireless channels among them, the nodes have access to a public noiseless channel to send side information in order to agree on a shared key\footnote{Please note that the public noiseless channel assumption could be relaxed, at the cost of using more sophisticated schemes and of a slight degradation of the key rate \cite{Cooperative}. Such an extension is left as future work.}. Furthermore, we assume Eve is several wavelengths apart from the legitimate nodes, which makes its channel independent of that between Alice and Bob. 
	The wireless channels are assumed to be affected by Rayleigh fading, but each channel changes in a time correlated manner, i.e., we assume that two consecutive channel gains, say $h_1$ and $h_2$ as shown in Fig. \ref{fig.2.b}, are correlated to each other.

	\subsection{Half Duplex}
	\label{ubSection: Half Duplex}
\begin{figure}[!t]
				\centering
				\subfigure[]{\includegraphics[width=0.48 \textwidth]{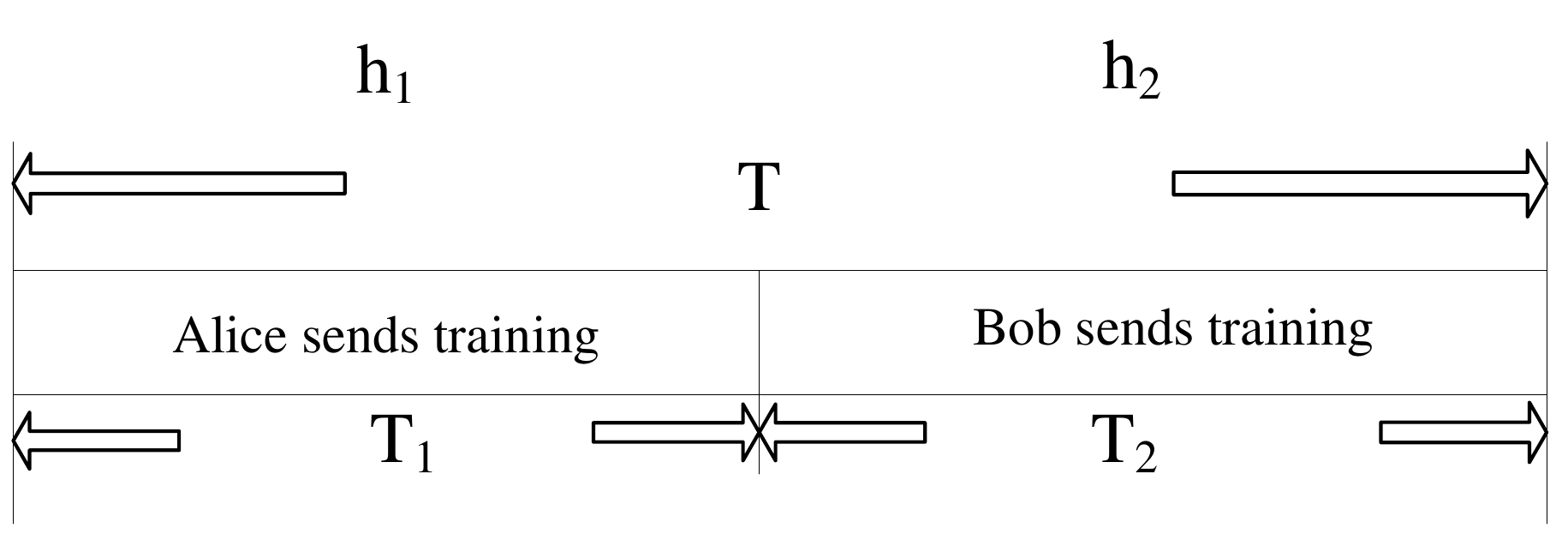}}
				\hfil
				\vspace{-0.35 cm}
				\subfigure[]{\includegraphics[width=0.48 
					\textwidth]{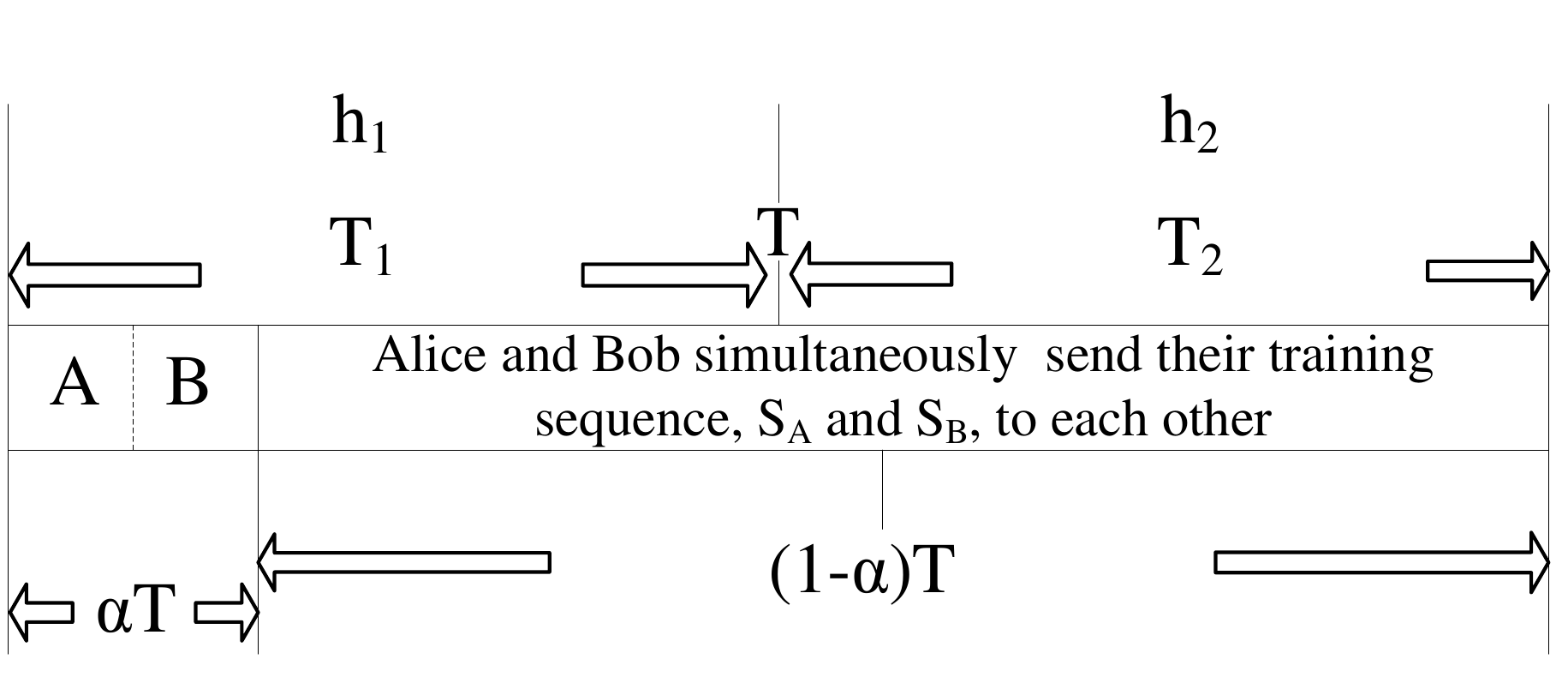}} 
				\caption{Time frame for the key agreement process. HD case (a), and FD case (b).}
				\label{fig.2.b}
			\end{figure}

	Fig. \ref{fig.2.b}a depicts the time frame for key generation process in the HD case. Each legitimate node sends training sequences in an intermittent manner to help the other party to estimate the wireless channel. 
	In the first time interval, $T_1$,  Alice transmits a sequence of training symbols, $\textbf{S}_{\text{A}}$, to Bob. Based on the received signal, Bob estimates the channel gain, ${{\tilde h}_{1,B}}$. In the remaining time interval, $T_2$, Bob transmits $\textbf{S}_{\text{B}}$ and Alice estimates ${{\tilde h}_{2,A}}$. The real Gaussian channel gains in these two time intervals are $h_1$ and $h_2$ with zero means and variances $\sigma_1^2$ and $\sigma_2^2$, respectively, which are in general the same. As explained, we assume that $h_1$ and $h_2$ are correlated with coefficient $\rho$. In addition, we assume that the channel gain over the interval that Alice or Bob senses remains fixed, i.e., $h_1$ and $h_2$ are fixed during intervals $T_1$ and $T_2$, respectively. Moreover, we assume that the times $T_1$ and $T_2$ indicate the number of symbols transmitted in the first and second time intervals.

	Thus, the received signal vectors at Alice and Bob are 
	\begin{align}
		& {{\bf{Y}}_A} = {h_2}{{\textbf{S}}_B} + {{\bf{N}}_A} \\
		& {{\bf{Y}}_B} = {h_1}{{\textbf{S}}_A} + {{\bf{N}}_B}
	\end{align}
	where ${{\bf{N}}_A}$ and ${{\bf{N}}_B}$ indicate the received noise vectors with iid zero mean Gaussian distributed elements with variance $\sigma^2$. Thus, the estimated channel gains at Alice and Bob become
	\begin{align}
		{{\tilde h}_{1,B}} = \frac{{{\bf{S}}_A^T}}{{{{\left\| {{{\bf{S}}_A}} \right\|}^2}}}{{\bf{Y}}_B} = {h_1} + \frac{{{\bf{S}}_A^T}}{{{{\left\| {{{\bf{S}}_A}} \right\|}^2}}}{{\bf{N}}_B} \\
		{{\tilde h}_{2,A}} = \frac{{{\bf{S}}_B^T}}{{{{\left\| {{{\bf{S}}_B}} \right\|}^2}}}{{\bf{Y}}_A} = {h_2} + \frac{{{\bf{S}}_B^T}}{{{{\left\| {{{\bf{S}}_B}} \right\|}^2}}}{{\bf{N}}_A}
	\end{align}
	
	During the time interval for key agreement, we assume that the nodes transmit with powers $P_A$ and $P_B$. It can be shown that $I({{\tilde h}_{1,B}};{{\tilde h}_{2,A}})=I({\bf{Y}}_B,{\bf{Y}}_A)$, so ${\bf{Y}}_B$ and ${\bf{Y}}_A$ are sufficient for key generation ($I(.;.)$ denotes mutual information). In addition, it is worth noting that ${{\tilde h}_{1,B}}$ and ${{\tilde h}_{2,A}}$ are correlated Gaussian random variables with variance $\sigma_1^2+\frac{\sigma^2}{{{{\left\| {{{\bf{S}}_A}} \right\|}^2}}}$ and $\sigma_2^2+\frac{\sigma^2}{{{{\left\| {{{\bf{S}}_B}} \right\|}^2}}}$ and correlation coefficient $\rho$.

	\subsection{Full-Duplex}
	\label{Subsection: Full-Duplex}
	Fig. \ref{fig.2.b}b depicts the time frame for key generation in the FD scenario. In the FD case the two legitimate nodes Alice and Bob are assumed to be imperfect FD. Note that in the IBFD case, we also consider the same setting where the channel takes two values of $h_1$ and $h_2$ over the intervals $T_1$ and $T_2$ in period $T$.  As discussed in Section \ref{Section:In-Band Full-Duplex Communication}, in the first stage of this diagram, i.e., the first $\alpha T$ symbol blocks, Alice and Bob estimate their own SI channel and also the analog and digital canceler parameters in order to enable any FD operation. In the second stage, thanks to the SI cancellation techniques initiated in the first stage, they simultaneously transmit training sequences over the shared wireless channel. This procedure is depicted in Fig.~3. In addition, it must be noted that, although in the first stage the nodes try to estimate their SI parameters, the other party can use the transmitted symbols over the air to estimate the shared channel. For instance, in the first ${{\left( {\alpha T} \right)} \mathord{\left/
			{\vphantom {{\left( {\alpha T} \right)} 2}} \right.
			\kern-\nulldelimiterspace} 2}$ period of the first stage, while Alice transmits symbols to initate its FD capability, Bob can hear the pilots as shown in Fig. 3a and can try to estimate the channel. The same happens when Bob transmits. 
	It is worth noting that due to the very slow changing feature of the SI channel compared to that of the shared wireless link between legitimate nodes \cite{DaurteExperimental} \cite{FDRadios}, there is no need to again estimate the SI channel parameters at the beginning of the $T_1$ interval.   
	
		\begin{figure}[!t]
			\centering
			\subfigure[First $\alpha T / 2$ training interval dedicated for Alice SI channel estimation phase.]{\includegraphics[width=0.25 \textwidth]{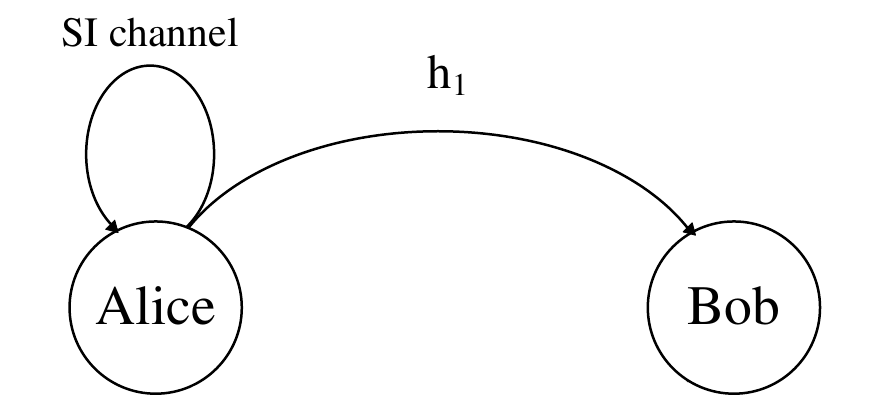}}
			\label{fig.3-a}
			\hfil
			\subfigure[Second $\alpha T / 2$ training interval dedicated for Bob SI channel estimation phase.]{\includegraphics[width=0.25 \textwidth]{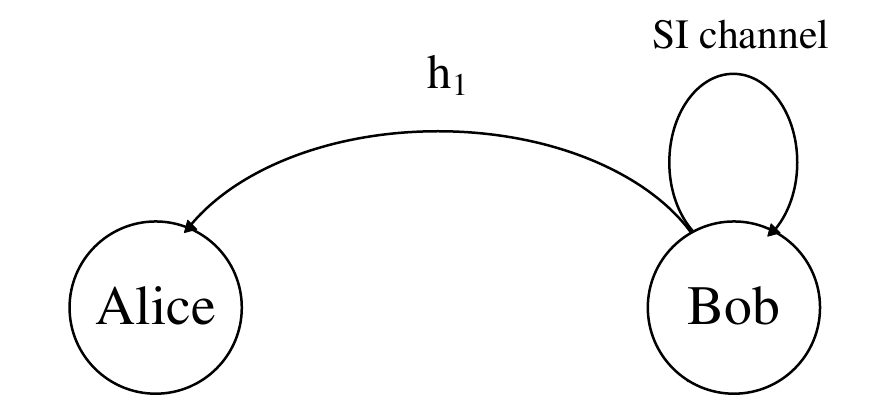}}
			\label{fig.3-b}
			\hfil
			\subfigure[The rest of the training interval dedicated for shared channel estimation phase.]{\includegraphics[width=0.25 \textwidth]{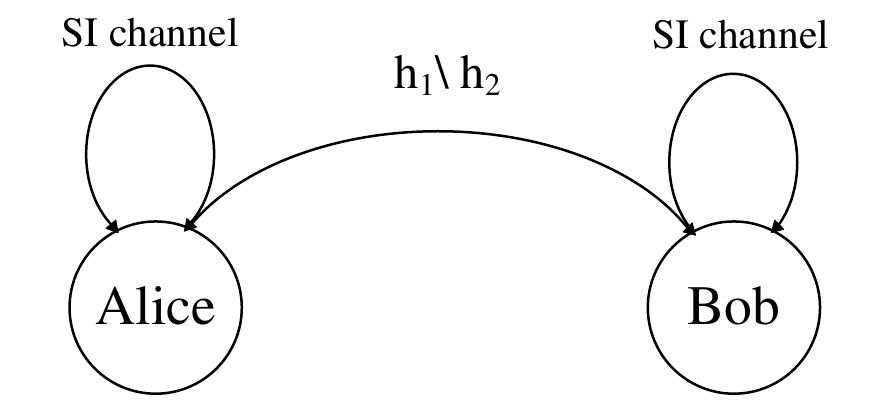}} 
			\label{fig.3-c}
			\hfill
			\caption{FD channel estimation steps.}
			\label{fig.3}
		\end{figure}
	
	During the first time interval $T_1$ the received signals at legitimate nodes can be represented as follows, where the superscript denotes the first time interval 
	\begin{align}
		&{\bf{Y}}_B^{\rm{FD},(1)} = {h_1}{\bf{S}}_A^{\rm{FD},(1)} + {\bf{RSI}}_B^{(1)} + {\bf{N}}_B^{\rm{FD},(1)}\\
		&{\bf{Y}}_A^{\rm{FD},(1)} = {h_1}{\bf{S}}_B^{\rm{FD},(1)} + {\bf{RSI}}_A^{(1)} + {\bf{N}}_A^{\rm{FD},(1)}
	\end{align} 
	and the estimated channels are
	\begin{align}
		{{\tilde h}_{1,B}} = \frac{{{{\left( {{\bf{S}}_A^{FD,(1)}} \right)}^T}}}{{{{\left\| {{\bf{S}}_A^{FD,(1)}} \right\|}^2}}}{\bf{Y}}_B^{FD,(1)} = {h_1} + \frac{{{{\left( {{\bf{S}}_A^{FD,(1)}} \right)}^T}}}{{{{\left\| {{\bf{S}}_A^{FD,(1)}} \right\|}^2}}}\left( {{\bf{RSI}}_B^{(1)} + {\bf{N}}_B^{FD,(1)}} \right)\\
		{{\tilde h}_{1,A}} = \frac{{{{\left( {{\bf{S}}_B^{FD,(1)}} \right)}^T}}}{{{{\left\| {{\bf{S}}_B^{FD,(1)}} \right\|}^2}}}{\bf{Y}}_A^{FD,(1)} = {h_1} + \frac{{{{\left( {{\bf{S}}_B^{FD,(1)}} \right)}^T}}}{{{{\left\| {{\bf{S}}_B^{FD,(1)}} \right\|}^2}}}\left( {{\bf{RSI}}_A^{(1)} + {\bf{N}}_A^{FD,(1)}} \right)
	\end{align}
	where ${\bf{Y}}^{\rm{FD},(1)}$, ${\bf{S}}^{\rm{FD},(1)}$, ${\bf{RSI}}^{(1)}$ and $\bf{N}^{\rm{FD},(1)}$ are the random variables in $T_1$. We must note that all of them are $(T_1-{{\alpha T} \mathord{\left/
			{\vphantom {{\alpha T} 2}} \right.
			\kern-\nulldelimiterspace} 2}) \times 1$ size vectors. In the ${{\alpha T} \mathord{\left/
			{\vphantom {{\alpha T} 2}} \right.
			\kern-\nulldelimiterspace} 2}$ portion of $T_1$ each node only estimates the SI parameters, including SI channel and analog and digital canceler coefficients. In the remaining portion $(T_1-{{\alpha T} \mathord{\left/
			{\vphantom {{\alpha T} 2}} \right.
			\kern-\nulldelimiterspace} 2})$ it will instead sense the shared channel.  In addition, the first ${{\alpha T} \mathord{\left/
			{\vphantom {{\alpha T} 2}} \right.
			\kern-\nulldelimiterspace} 2}$ elements of the RSI vector are zero, since in the first ${{\alpha T} \mathord{\left/
			{\vphantom {{\alpha T} 2}} \right.
			\kern-\nulldelimiterspace} 2}$ symbol blocks of $(T_1-{{\alpha T} \mathord{\left/
			{\vphantom {{\alpha T} 2}} \right.
			\kern-\nulldelimiterspace} 2})$, the nodes are operating in HD mode, i.e., they are only sensing, which does not cause any RSI to the nodes.
	
	In the remaining $T_2$ symbol blocks, the received and estimated channels are respectively
	\begin{align}
		{\bf{Y}}_B^{FD,(2)} = {h_2}{\bf{S}}_A^{FD,(2)} + {\bf{RSI}}_B^{(2)} + {\bf{N}}_B^{FD,(2)}\\
		{\bf{Y}}_A^{FD,(2)} = {h_2}{\bf{S}}_B^{FD,(2)} + {\bf{RSI}}_A^{(2)} + {\bf{N}}_A^{FD,(2)}
	\end{align}
	and
	\begin{align}
		{{\tilde h}_{2,B}} = \frac{{{{\left( {{\bf{S}}_A^{FD,(2)}} \right)}^T}}}{{{{\left\| {{\bf{S}}_A^{FD,(2)}} \right\|}^2}}}{\bf{Y}}_B^{FD,(2)} = {h_2} + \frac{{{{\left( {{\bf{S}}_A^{FD,(2)}} \right)}^T}}}{{{{\left\| {{\bf{S}}_A^{FD,(2)}} \right\|}^2}}}\left( {{\bf{RSI}}_B^{(2)} + {\bf{N}}_B^{FD,(2)}} \right) \\
		{{\tilde h}_{2,A}} = \frac{{{{\left( {{\bf{S}}_B^{FD,(2)}} \right)}^T}}}{{{{\left\| {{\bf{S}}_B^{FD,(2)}} \right\|}^2}}}{\bf{Y}}_A^{FD,(2)} = {h_2} + \frac{{{{\left( {{\bf{S}}_B^{FD,(2)}} \right)}^T}}}{{{{\left\| {{\bf{S}}_B^{FD,(2)}} \right\|}^2}}}\left( {{\bf{RSI}}_A^{(2)} + {\bf{N}}_A^{FD,(2)}} \right)
	\end{align}
	where the vector sizes during the $T_2$ symbol blocks are $T_2 \times 1$. In addition, it must be noted that the estimated channels in these periods at the legitimate nodes are noisy versions of each other.

	\begin{figure}[!t]
		\centering
		\includegraphics[width=0.27 \textwidth]{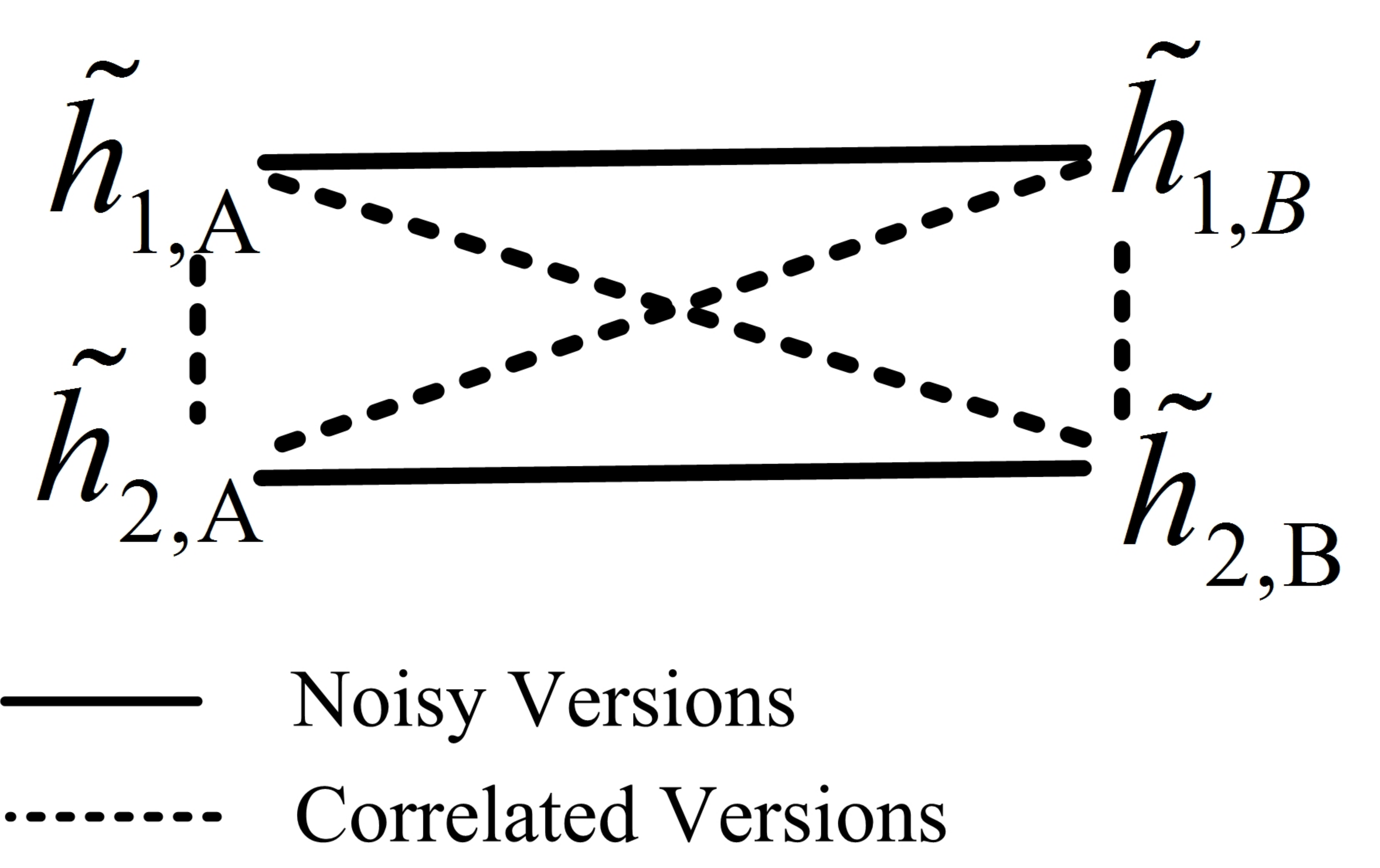}
		\caption{Alice and Bob channel estimations}
		\label{Fig4}
	\end{figure}
	
	\section{Secret Key Rate}
	\label{Secrecy Key Generation}
	After channel estimations, the two legitimate nodes have a common source of randomness to generate secure keys. As discussed in \cite{Maurer}, the upper bound for the key generation rate is equal to the conditional mutual information between the estimated channel vectors ${{\tilde {\bf {h}}}_{A}}$ and  ${{\tilde {\bf {h}}}_{B}}$ at the legitimate nodes of the shared wireless link given Eve's estimations, i.e., ${{\tilde {\bf{h}}}_{E}}$, as follows
	\begin{align}
		{R_{{\textrm{Key}}}} &= I\left( {\left. {{{\tilde {\bf{h}}}_{A}};{{\tilde {\bf{h}}}_{B}}} \right|{{\tilde {\bf{h}}}_{E}}} \right) \\
		&\mathop  = \limits^{(a)}  I\left( {{{\tilde {\bf{h}}}_{A}};{{\tilde {\bf{h}}}_{B}}} \right)
		\label{DefinitionofKeyRate}
	\end{align}
	in this formula (a) follows from the distance assumption between Eve and the legitimate nodes which makes Eve's channel estimates independent of the legitimate nodes'.
	
	For key agreement, the legitimate nodes exploit Slepian-Wolf source coding in the information reconciliation step. In fact, due to the estimation errors the two nodes cannot agree on a single key in one time slot, consequently one of them should transmit side information through a noiseless channel to help the other party in order to eliminate any mismatch in the final generated key \cite{Cooperative}. This noiseless channel, which can be heard by every node in the network, including Eve, is known as public channel.   
	In the information reconciliation stage, Alice (Bob) transmits $H({\bf{Y}}_A|{\bf{Y}}_B)$ (or $H({\bf{Y}}_B|{\bf{Y}}_A)$) bits of information using Slepian-Wolf coding through the public channel. These bits, although observable by Eve, do not leak any information about the key. 
	\subsection{Half Duplex Shared Key Rate}
	\label{Subsection:Half Duplex Shared Key Rate}
	In the HD case, Alice and Bob have two correlated estimations of the channel as ${{\tilde {\bf {h}}}_{A}} = \tilde{h}_{{2,A}}$ and ${{\tilde {\bf {h}}}_{B}} = {\tilde h}_{1,B}$.  Thus based on \eqref{DefinitionofKeyRate} it can be shown that the key rate would be
	\begin{align}
		\nonumber
		{R_{\textrm{Key,HD}}} &= \frac{1}{T}I\left( {{{\tilde h}_{1,B}};{{\tilde h}_{2,A}}} \right)\\
		&= \frac{1}{{2T}}\log \left[ {\frac{{\left( {\sigma _1^2 + \frac{{{\sigma ^2}}}{{{P_A}{T_1}}}} \right)\left( {\sigma _2^2 + \frac{{{\sigma ^2}}}{{{P_B}{T_2}}}} \right)}}{{\left( {\sigma _1^2 + \frac{{{\sigma ^2}}}{{{P_A}{T_1}}}} \right)\left( {\sigma _2^2 + \frac{{{\sigma ^2}}}{{{P_B}{T_2}}}} \right) - {\rho ^2}\sigma _1^2\sigma _2^2}}} \right]
		\label{HDKEyrate}
	\end{align}
	Where $\rho$ indicates the correlation coefficient between the two estimated channels at nodes $A$ and $B$, namely $ {{{\tilde h}_{1,B}}}$ and ${{{\tilde h}_{2,A}}}$.   
	In an ideal case of channel estimations, where the nodes estimate the channel simultaneously, we would have $\rho =1$ and the result would become the same as indicated in \cite{Cooperative}.

	\subsection{In-Band Full-Duplex Shared Key Rate}
	\label{Subsection:In-Band Full-Duplex Shared Key Rate}
	In the FD case, Alice would have its own observation of the shared channel as a vector ${{\tilde {\bf{h}}}_{A}} = \left( {{{\tilde h}_{1,A}},{{\tilde h}_{2,A}}} \right)$ and Bob would have $ {{\tilde {\bf{h}}}_{B}} = \left( {{{\tilde h}_{1,B}},{{\tilde h}_{2,B}}} \right)$. Based on the system model ${{\tilde h}_{i,A}} \textrm{ and }{{\tilde h}_{i,B}}$ are noisy versions, while ${{\tilde h}_{i,A \setminus B}} \textrm{  and } {{\tilde h}_{\overline i ,\left( {A \setminus B} \right)}}$ are correlated versions of each other. Here, $i \in \{1,2\}$ and $\overline i$ is the complement of $i$. The diagram showing these relations between estimated channels at two nodes is shown in Fig. \ref{Fig4}, consequently, the key rate generation in this scheme according to \eqref{DefinitionofKeyRate} is
	\begin{equation}
		{R_{\textrm{{Key,FD}}}} = \frac{1}{T}I\left( {{{\tilde h}_{1,A}},{{\tilde h}_{2,A}};{{\tilde h}_{1,B}},{{\tilde h}_{2,B}}} \right).
	\end{equation}
	
	In the symmetric scenario, i.e., $P_A = P_B =P$, $T_1 = T_2$, the key rate is computed as in (24), where $T = T_1 + T_2$. The proof is provided in Appendix B.

	\section{Numerical Results and Analysis}
	\label{Section: Simulation Results and Analysis}
	To investigate the impact of FD operation on the secure key rate generation via fading channels we conducted a few numerical simulations. In the simulations, the transmit powers and RSI variances are normalized to the noise level, i.e., $\sigma^2 = 1$, and for a fair comparison, the transmit power of the nodes in FD is half of that in HD.  
	
%	$\sigma_{A,FD}^2=  \sigma_{RSI,A}^2 + \sigma^2 = \sigma_{B,FD}^2 = \sigma_{RSI,B}^2 + \sigma^2 = 0$

\begin{figure}[!t]
		\centering
		\includegraphics[width=0.45 \textwidth]{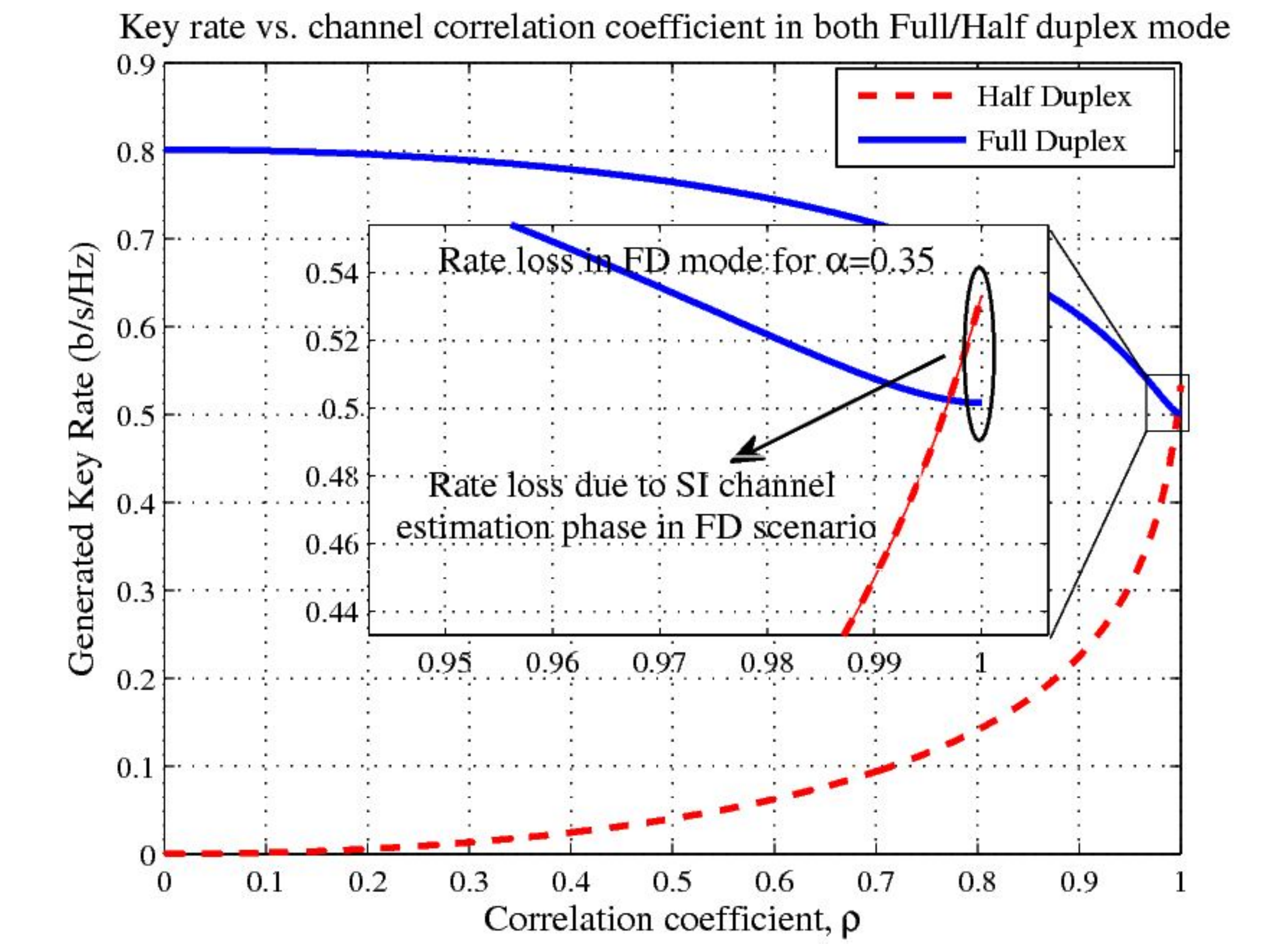}
		\caption{Key rate versus estimated channels' correlation coefficient}
		\label{Result1}
	\end{figure}

	In Fig. \ref{Result1} we compare the key rate in both FD and HD cases as a function of the correlation coefficient of the estimated channels. We set $T_1 = T_2 = 2.5$ seconds, $\textrm{SNR}_A = \frac{{{P_A}}}{{{\sigma ^2}}} = \textrm{SNR}_B = \frac{{{P_B}}}{{{\sigma ^2}}} = 10$ dB, $\sigma_{RSI,A}^2 = \sigma_{RSI,B}^2 = 0$ (perfect FD), $\alpha=0.35$, and $\sigma_1^2 = \sigma_2^2 = 5$ dB. It can be seen from the figure that for small values of the correlation coefficient ($\rho \rightarrow 0$) the HD key rate decreases dramatically, while the FD key rate not only does not decrease but even increases. This is due to the fact that in the less correlated scenarios the FD devices sense a more sharply changing channel, and consequently can exploit this additional random shared source to generate the key while in the HD case the less correlated channels mean losing the shared random source for key generation purposes.  Moreover, in the very highly correlated scenarios, $\rho \rightarrow 1$, the FD case will lose key rate compared to HD, since in this case
	FD devices need to estimate SI channel parameters ($\alpha = 0.35$) while the random sources available at both FD and HD devices are the same.
	
	\begin{figure}[!t]
		\centering
		\includegraphics[width=.45 \textwidth]{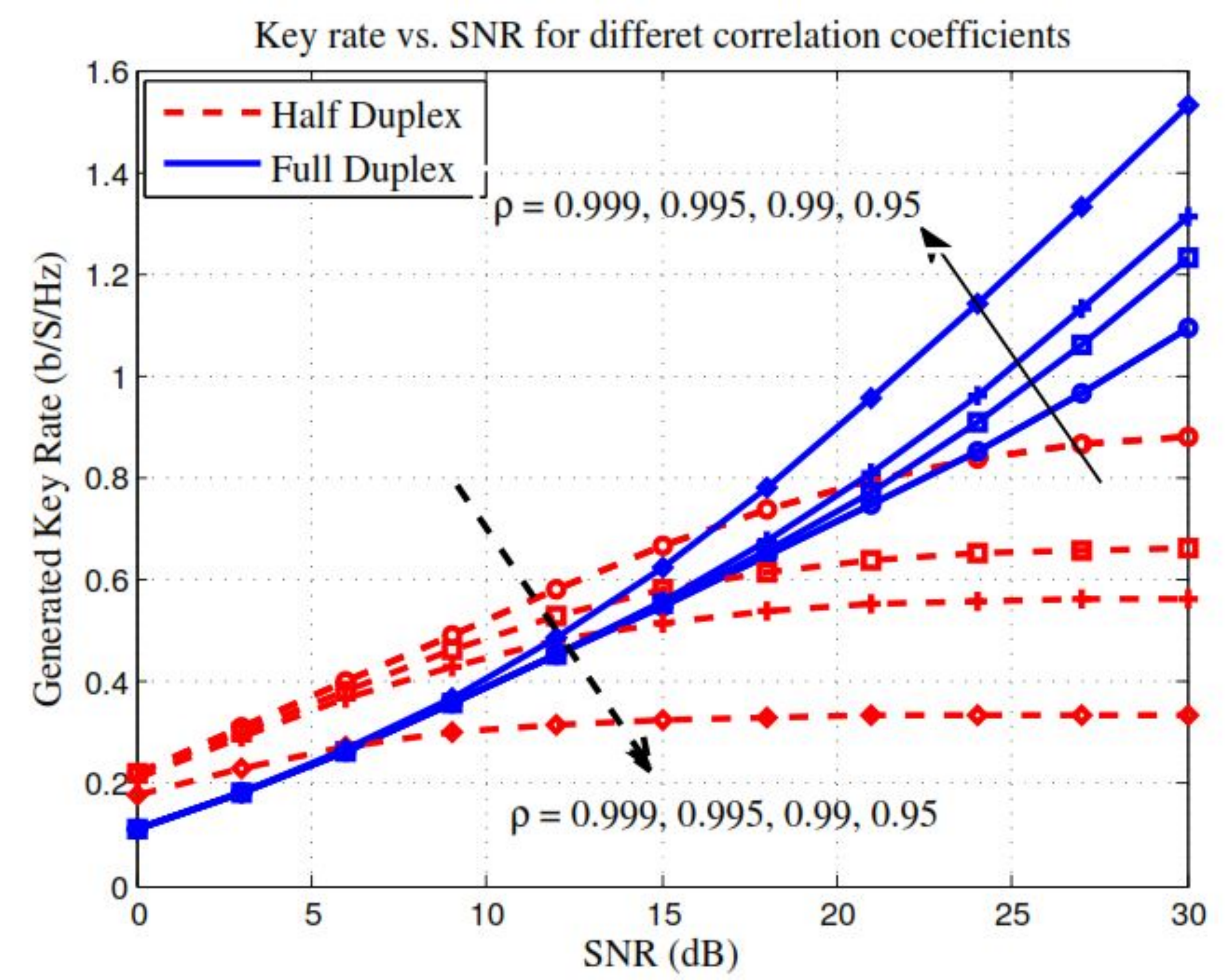}
		\vspace{-.5cm}
		\caption{Key rate versus SNR for different correlation coefficients}
		\label{Result5}
	\end{figure}
	
	In Fig. \ref{Result5} we plotted the key rate as a function of the transmission power, for different correlation coefficient in both FD and HD cases. We set $\sigma_{A,FD}^2 = \sigma_{B,FD}^2 = \sigma^2 =0$ dB, i.e., the RSI signal is as strong as the noise. From the figure, it can be seen that in high SNR regime the HD key rate tends to a constant which, based on \eqref{HDKEyrate}, is $\frac{1}{T}\log (\frac{1}{{\sqrt {1 - {\rho ^2}} }})$. Evidently, for $\rho \ne 1$, which is the case in HD mode, this would be the main limitation of secrecy key generation. However, for the FD case, as shown, the higher the SNR the better the key rate. In addition, similar to Fig. \ref{Result1}, the less the correlation coefficient the higher the key rate. 
	
	It is worth noting that even in the FD case, if the SNR increases dramatically the key rate will tend to a constant. This is due to the fact that the RSI channel variance, $\sigma_{RSI}^2$, and consequently also $\sigma_{FD}^2$, are functions of the transmit power of the FD device, thus if $P \rightarrow \infty$ then $\sigma_{FD}^2 \rightarrow \infty$, which leads to a constant ratio, i.e., $\frac{\sigma_{FD}^2}{P} \rightarrow \eta$, and consequently would cause  the FD key rate to approach a constant as $P \rightarrow \infty$. However for reasonable SI cancellation, as reported in Fig. \ref{Result5} in high SNRs, the better gain is expected in the FD case.
	
	Finally, as seen from Fig. 6, in low SNRs, and highly correlated channel observations (ideal case of reciprocal channel measurements, $\rho \rightarrow 1$), the FD key rate is less than the HD counterpart. This rate loss is due to the RSI and SI channel estimation phase overheads imposed to the FD systems, while the randomness sources available in both FD and HD cases are the same.
	
	\begin{figure*}[!t]
		% ensure that we have normalsize text
		% Store the current equation number.
		\setcounter{MYtempeqncnt}{\value{equation}}
		% Set the equation number to one less than the one
		% desired for the first equation here.
		% The value here will have to changed if equations
		% are added or removed prior to the place these
		% equations are referenced in the main text.
		\setcounter{equation}{23}
		\begin{equation}
			\label{eqn_dbl_y}
			{R_{\textrm{{Key,FD}}}} = \frac{1}{{2T}} \times \log \left[ {\frac{{\left( {\left( {\sigma _1^2 + \frac{{2\sigma _{A,FD}^2}}{{P\left( {1 - \alpha } \right)T}}} \right)\left( {\sigma _2^2 + \frac{{2\sigma _{A,FD}^2}}{{P\left( {1 - \alpha } \right)T}}} \right) - {\rho ^2}\sigma _1^2\sigma _2^2} \right)\left( {\left( {\sigma _1^2 + \frac{{2\sigma _{B,FD}^2}}{{P\left( {1 - \alpha } \right)T}}} \right)\left( {\sigma _2^2 + \frac{{2\sigma _{B,FD}^2}}{{P\left( {1 - \alpha } \right)T}}} \right) - {\rho ^2}\sigma _1^2\sigma _2^2} \right)}}{{{{\left( {\frac{{2\sigma _{B,FD}^2}}{{P\left( {1 - \alpha } \right)T}}} \right)}^2}\left\{ {\left[ {\sigma _1^2\left( {1 + \frac{{\sigma _{A,FD}^2}}{{\sigma _{B,FD}^2}}} \right) + \frac{{2\sigma _{A,FD}^2}}{{P\left( {1 - \alpha } \right)T}}} \right]\left[ {\sigma _2^2\left( {1 + \frac{{\sigma _{A,FD}^2}}{{\sigma _{B,FD}^2}}} \right) + \frac{{2\sigma _{A,FD}^2}}{{P\left( {1 - \alpha } \right)T}}} \right] - {\rho ^2}\sigma _1^2\sigma _2^2{{\left( {1 + \frac{{\sigma _{A,FD}^2}}{{\sigma _{B,FD}^2}}} \right)}^2}} \right\}}}} \right]
		\end{equation}
		\begin{align}
			{\bf{C = }}\left[ {\begin{array}{*{20}{c}}
					{\sigma _1^2 + \frac{{\sigma _{RS{I_A}}^2}}{{{P_B}\left( {{T_1} - 2\alpha T} \right)}} + \frac{{{\sigma ^2}}}{{{P_B}\left( {{T_1} - \alpha T} \right)}}}&{\rho {\sigma _1}{\sigma _2}}&{\sigma _1^2}&{\rho {\sigma _1}{\sigma _2}}\\
					{\rho {\sigma _1}{\sigma _2}}&{\sigma _2^2 + \frac{{\sigma _{RS{I_A}}^2 + {\sigma ^2}}}{{{P_B}{T_2}}}}&{\rho {\sigma _1}{\sigma _2}}&{\sigma _2^2}\\
					{\sigma _1^2}&{\rho {\sigma _1}{\sigma _2}}&{\sigma _1^2 + \frac{{\sigma _{RS{I_B}}^2}}{{{P_A}\left( {{T_1} - 2\alpha T} \right)}} + \frac{{{\sigma ^2}}}{{{P_A}\left( {{T_1} - \alpha T} \right)}}}&{\rho {\sigma _1}{\sigma _2}}\\
					{\rho {\sigma _1}{\sigma _2}}&{\sigma _2^2}&{\rho {\sigma _1}{\sigma _2}}&{\sigma _2^2 + \frac{{\sigma _{RS{I_B}}^2 + {\sigma ^2}}}{{{P_A}{T_2}}}}
				\end{array}} \right]
			\end{align}
			\begin{equation}
				\left| {{\bf{C}} } \right| = {\left( {\frac{{2\sigma _{B,FD}^2}}{{P\left( {1 - \alpha } \right)T}}} \right)^2} \times \left[ {\left( {\sigma _1^2\left( {1 + \frac{{\sigma _{A,FD}^2}}{{\sigma _{B,FD}^2}}} \right) + \frac{{2\sigma _{A,FD}^2}}{{P\left( {1 - \alpha } \right)T}}} \right)\left( {\sigma _2^2\left( {1 + \frac{{\sigma _{A,FD}^2}}{{\sigma _{B,FD}^2}}} \right) + \frac{{2\sigma _{A,FD}^2}}{{P\left( {1 - \alpha } \right)T}}} \right) - {\rho ^2}\sigma _1^2\sigma _2^2{{\left( {1 + \frac{{\sigma _{A,FD}^2}}{{\sigma _{B,FD}^2}}} \right)}^2}} \right]
			\end{equation}
			% Restore the current equation number.
			\setcounter{equation}{\value{MYtempeqncnt}+1}
			% IEEE uses as a separator
			\hrulefill
			% The spacer can be tweaked to stop underfull vboxes.
		\end{figure*}
		
	\section{Conclusion}
	\label{Section: Conclusion}
	In this paper we investigated the impact of IBFD wireless communication on the performance of secret key generation. A simple framework to generate keys based on FD devices is proposed and is shown to enhance the key rate compared to its HD counterpart by considering the non-reciprocity inherent in channel measurements. In addition, in high SNRs, it is shown that the HD rate tends to a constant when the legitimate nodes have correlated observations, while for the FD case it would tend to a rate which is a function of the SI cancellation performance. However, in reasonable SI cancellation performance, i.e., when the RSI and the noise are comparable, the FD key rate is much higher than that of HD. The high performance of the FD case is due to the fact that although FD devices should tolerate some overheads like RSI and the SI channel estimation phase, they would have both correlated and noisy versions of the estimated channels, which will provide an additional source of randomness to generate the key.
	
	A possible extension of this work would be to explore how the individual time-frames in the proposed scheme for self-interference channel estimation and direct channel estimation may be optimized in light of the associated practical requirements. Other directions of interest include considering more sophisticated schemes to enable key agreement without the need for a noiseless public channel and/or defining imperfect FD relays to achieve multiplexing gain for the generated key in line with what is proposed in \cite{Cooperative}.

		\appendix
		
		\section{}
		\label{Appendix B}
		Define the estimated channel vectors at Alice and Bob as
		\begin{align}
			{{\bf{\tilde h}}_A} = \left( {\begin{array}{*{20}{c}}
					{{{\tilde h}_{1,A}}}\\
					{{{\tilde h}_{2,A}}}
				\end{array}} \right),{\mkern 1mu} {{\bf{\tilde h}}_B} = \left( {\begin{array}{*{20}{c}}
				{{{\tilde h}_{1,B}}}\\
				{{{\tilde h}_{2,B}}}
			\end{array}} \right)
		\end{align}
		
		The key rate in this case is ${R_{\textrm{{Key,FD}}}} = \frac{1}{T}I\left( {{{{\bf{\tilde h}}}_A};{{{\bf{\tilde h}}}_B}} \right)$. Since ${{{{\bf{\tilde h}}}_A}}$ and ${{{{\bf{\tilde h}}}_B}}$ are multivariate Gaussian distributed, the mutual information can be found as follows \cite{Entropyinfotheory}
		\begin{equation}
			{R_{\textrm{{Key,FD}}}} = \frac{1}{T}I\left( {{{{\bf{\tilde h}}}_A};{{{\bf{\tilde h}}}_B}} \right) = \frac{1}{{2T}}\log \left[ {\frac{{\left| {{{\bf{C}}_{{{{\bf{\tilde h}}}_A}{\bf{,}}{{{\bf{\tilde h}}}_A}}}} \right|\left| {{{\bf{C}}_{{{{\bf{\tilde h}}}_B}{\rm{,}}{{{\bf{\tilde h}}}_B}}}} \right|}}{{\left| {\bf{C}} \right|}}} \right]
			\label{FDKeyrate}
		\end{equation}
		where ${{{\bf{C}}_{{{{\bf{\tilde h}}}_i}{\rm{,}}{{{\bf{\tilde h}}}_i}}}}$ for $i \in \left\{ {A,B} \right\}$ indicates the covariance matrix of ${{\bf{\tilde h}}}_i$, and  
		${\bf{C = }}\left[ {\begin{array}{*{20}{c}}
			{{{\bf{C}}_{{{{\bf{\tilde h}}}_A}{\bf{,}}{{{\bf{\tilde h}}}_A}}}}&{{{\bf{C}}_{{{{\bf{\tilde h}}}_A}{\rm{,}}{{{\bf{\tilde h}}}_B}}}}\\
			{{{\bf{C}}_{{{{\bf{\tilde h}}}_B}{\bf{,}}{{{\bf{\tilde h}}}_A}}}}&{{{\bf{C}}_{{{{\bf{\tilde h}}}_B}{\rm{,}}{{{\bf{\tilde h}}}_B}}}}
			\end{array}} \right]$. In \eqref{FDKeyrate}, $\left| . \right|$ indicates the determinant operator. According to the estimated channel gains in FD scenario, provided in Section \ref{Subsection: Full-Duplex}, the covariance matrices are as follows
		
		\begin{align}
			{{\bf{C}}_{{{{\bf{\tilde h}}}_A}{\rm{,}}{{{\bf{\tilde h}}}_A}}} = \left[ {\begin{array}{*{20}{c}}
					{\sigma _1^2 + \frac{{\sigma _{RS{I_A}}^2}}{{{P_B}\left( {{T_1} - 2\alpha T} \right)}} + \frac{{{\sigma ^2}}}{{{P_B}\left( {{T_1} - \alpha T} \right)}}}&{\rho {\sigma _1}{\sigma _2}}\\
					{\rho {\sigma _1}{\sigma _2}}&{\sigma _2^2 + \frac{{\sigma _{RS{I_A}}^2 + {\sigma ^2}}}{{{P_B}{T_2}}}}
				\end{array}} \right] 
				\label{AA}
				\\ 
				{{\bf{C}}_{{{{\bf{\tilde h}}}_B}{\rm{,}}{{{\bf{\tilde h}}}_B}}} = \left[ {\begin{array}{*{20}{c}}
						{\sigma _1^2 + \frac{{\sigma _{RS{I_B}}^2}}{{{P_A}\left( {{T_1} - 2\alpha T} \right)}} + \frac{{{\sigma ^2}}}{{{P_A}\left( {{T_1} - \alpha T} \right)}}}&{\rho {\sigma _1}{\sigma _2}}\\
						{\rho {\sigma _1}{\sigma _2}}&{\sigma _2^2 + \frac{{\sigma _{RS{I_B}}^2 + {\sigma ^2}}}{{{P_A}{T_2}}}}
					\end{array}} \right]
					\label{BB}
				\end{align}
				\begin{equation}
					{{\bf{C}}_{{{{\bf{\tilde h}}}_A}{\bf{,}}{{{\bf{\tilde h}}}_B}}} = {\bf{C}}_{{{{\bf{\tilde h}}}_B}{\rm{,}}{{{\bf{\tilde h}}}_A}}^T = \left[ {\begin{array}{*{20}{c}}
							{\sigma _1^2}&{\rho {\sigma _1}{\sigma _2}}\\
							{\rho {\sigma _1}{\sigma _2}}&{\sigma _2^2}
						\end{array}} \right]
					\end{equation} 
					
					The key rate in the FD case, $R_{\textrm{{Key,FD}}}$, can be found by replacing the determinant of matrices \eqref{AA}, \eqref{BB}, and (25) into \eqref{FDKeyrate}. Due to the symmetric matrix $\bf{C}$, by doing algebraic row and column operations the determinant can be found more easily as in (26). Thus, the key rate in FD case is as in (24), where in this relation $\sigma _{FD}^2 = \sigma _{RSI}^2 + {\sigma ^2}$.

					\bibliographystyle{IEEEtran}
					\bibliography{References}

					% that's all folks
				\end{document}